\documentclass[letter,bibyear]{aa} 

\usepackage{graphicx}
\usepackage{txfonts}
\usepackage{hyperref}
\def\txs0506{\object{TXS\,0506+056}}

\begin{document}

   \title{Apparent superluminal core expansion  and limb brightening in the candidate neutrino blazar \txs0506}

   \author{%
                E. Ros \inst{1}
          \and
                M. Kadler\inst{2}
          \and
        M. Perucho\inst{3,4}
           \and
        B. Boccardi\inst{1}
           \and        
        H.-M. Cao\inst{5}  
        \and
        M. Giroletti\inst{5}
        \and
                F. Krau\ss{}\inst{6}
                \and
                R. Ojha\inst{7,8,9}
          }

   \institute{
                Max-Planck-Institut f\"ur Radioastronomie, 
                Auf dem H\"ugel 69, D-53121 Bonn, Germany\\
        \email{ros@mpifr-bonn.mpg.de}
          \and
                Lehrstuhl für Astronomie, Universität Würzburg, Emil-Fischer-Straße 31, D-97074 Würzburg, Germany
                \and
                Departament d'Astronomia i Astrof\'{\i}sica, 
                Universitat de Val\`encia, c/ Dr. Moliner 50, E-46100 Burjassot, Val\`encia, Spain
                \and
                Observatori Astron\`omic,
                Universitat de Val\`encia, c/ Catedr\`atic Jos\'e Beltr\'an Mart\'{\i}nez 2, E-46980 Paterna, Val\`encia, Spain
         \and
         INAF – Istituto di Radioastronomia, Via Gobetti 101, I-40129, Bologna, Italy
        \and
          Department of Astronomy and Astrophysics, Pennsylvania State University, University Park, PA 16801, USA
          \and
          National Aeronautics and Space Administration/Goddard Space Flight Center, 
                Greenbelt, MD 20771, USA
                \and 
                University of Maryland, Baltimore County, 1000 Hilltop Cir, Baltimore, MD, 21250 USA
                \and
                Catholic University of America, Washington, DC, 20064, USA
             }

   \date{Submitted: November 28, 2019; 
   Accepted: December 3, 2019}

 
  \abstract
{IceCube has reported a very-high-energy neutrino (IceCube-170922A) in a region containing the blazar TXS\,0506+056. Correlated gamma-ray activity has led to the first high-probability association of a high-energy neutrino with an extragalactic source. This blazar has been found to be in a radio outburst during the neutrino event.} 
{%
Our goal is to probe the sub-milliarcsecond properties of the radio jet right after the neutrino detection and during the further evolution of the radio outburst. 
}
{%
We performed target of opportunity observations 
at {43\,GHz frequency} using very long baseline interferometry imaging, corresponding to 7\,mm in wavelength,
with the Very Long Baseline Array two and eight months after the neutrino event.
}
   {We produced two images of the radio jet of TXS\,0506+056 at 43\,GHz with angular resolutions of $(0.2 \times 1.1)$\,mas and $(0.2 \times 0.5)$\,mas, respectively. The source shows a compact, high brightness temperature core, albeit not approaching the equipartition limit 
and a bright and originally very collimated inner jet.
Beyond approximately 0.5\,mas from the millimeter-VLBI core, the jet loses this tight collimation and expands rapidly. 
During the months after the neutrino
   event associated with this source, the overall flux density is rising. This flux {density} increase happens solely within the core. Notably, the core expands in size with apparent superluminal velocity during these six months so that the brightness temperature drops by a factor of three despite the strong flux {density} increase.}
   {The radio jet of TXS\,0506+056 shows strong signs of deceleration and/or a spine-sheath structure within the inner $1$\,mas, corresponding to about $70$\,pc to 140\,pc in deprojected distance, from the millimeter-VLBI core. This structure is consistent with theoretical models that attribute the neutrino and gamma-ray production in TXS\,0506+056 to interactions of electrons and protons in the highly relativistic jet spine with external photons originating from a slower moving jet region. Proton loading due to jet-star interactions in the inner host galaxy is suggested as the possible cause of deceleration.
}


 \maketitle
%
\section{Introduction}
On September 22, 2017, the IceCube Neutrino Observatory (IceCube Collaboration \cite{ice17})
detected a $\sim$290\,TeV neutrino (IceCube-170922A) from a direction consistent with the flaring $\gamma$-ray, radio-loud blazar \txs0506, with a positional uncertainty of $0.4^\circ$ to $0.8^\circ$ (IceCube Collaboration et al. \cite{ice18a}), resulting in a $\sim3\sigma$ significance of an association. 
In addition, the IceCube Collaboration (\cite{ice18b}) report a $\sim3.5\sigma$ excess of $13 \pm 5$ neutrino events in the direction of \txs0506 during a six-month period in 2014-2015 without accompanying bright $\gamma$-ray flaring. 
These two observations diverge
in that Reimer et al.~(\cite{reimer19}) and Rodrigues et al.~(\cite{rodrigues2019}) find that single-zone models cannot explain the {spectral energy distribution} (SED) during the $\gamma$-ray faint 2014--2015 period.

\txs0506\ has a redshift of 0.3365$\pm$0.0010. This corresponds to a luminosity distance of 1762\,Mpc. An angular distance of 1\,mas in the sky corresponds to a linear scale of 4.78\,pc\footnote{We use $H_0 = 71 \mathrm{km}\,\mathrm{s}^{-1}\,\mathrm{Mpc}^{-1}$, $\Omega_\Lambda = 0.73$ and $\Omega_m = 0.27$.}. This source is classified as a BL\,Lac object, although Padovani et al. (\cite{pad19}) present arguments for a quasar-like nature for this source.

An intensive multiwavelength campaign followed the IceCube detection, covering the whole electromagnetic spectrum; see IceCube et al.\ (\cite{ice18a}). These observations resulted in the first redshift measurement (Paiano et al. \cite{paiano18}) and the detection of \txs0506\ at teraelectronvolt energies by the MAGIC telescopes (Ansoldi et al.~\cite{magic2018}).
In the radio band,
the Owens Valley Radio Observatory has been monitoring this source over the last ten years at 15\,GHz (Richards et al.\ \cite{ric11}) and has registered an increase in the flux density since early 2017\footnote{See the corresponding light curve at their 
\href{http://www.astro.caltech.edu/ovroblazars/data.php?page=data_return&source=J0509+0541}{webpage}} .  
The MOJAVE program ({Monitoring of Jets in Active Galactic Nuclei with VLBA Experiments}, Lister et al.\ \cite{lis09,lis18}) is also monitoring this source with 15\,GHz very long baseline interferometry (VLBI) imaging since 2009\footnote{See the imaging database at the \href{https://www.physics.purdue.edu/MOJAVE/sourcepages/0506+056.shtml}{MOJAVE webpage}.}.  The MOJAVE collaboration reports maximum jet speeds of 0.98$\pm$0.31\,c (Lister et al.~\cite{lis19}), which is a relatively low apparent jet speed compared to the overall speed distribution in the MOJAVE program. Such low jet velocities, however, are frequently reported for a number of BL\,Lac objects emitting tera electron volts (e.g., Piner \& Edwards (\cite{piner14,piner18}), and references therein).

Based on public archival MOJAVE data,
Kun et al.~(\cite{kun18}) report that the increase in flux density observed can be assigned to the most central region of the source, associated with the 2\,cm core of the parsec-scale image (i.e., with the inner $\sim 1$\,mas).  
Another publication making use of the public MOJAVE database (Britzen et al.~\cite{bri19}) claims the possible presence of a second jet in \txs0506, which might interact with the primary jet giving rise to enhanced neutrino emission. 
Higher frequency VLBI observations provide increased angular resolution and allow us to peer deeper into the VLBI core where the radio variability originates. These observations also allow us to test for the presence of a possible secondary jet core and for signs of a jet-jet interaction as proposed by Britzen et al.~(\cite{bri19}). Moreover,
high-resolution VLBI observations can probe the internal jet structure
and test the existence of velocity gradients, either along the jet axis (jet acceleration/deceleration) 
or perpendicular to it (spine-sheath structure).
Such gradients might play a crucial role in the $\gamma$-ray 
emission and correlated neutrino production,
as proposed by several authors in relation to \txs0506\ (Tavecchio et al.~\cite{tav14}; Righi et al.~\cite{rig17}; Ansoldi et al.~\cite{magic2018}; Zhang et al. \cite{zhang19}).
We present millimeter-VLBI observations of \txs0506\ during the radio outburst associated with the IceCube-170922A neutrino event.

\section{Observations and data reduction}
We observed \txs0506\ using the VLBA as soon as possible after the
IceCube detection.  At that time, the easternmost
station, Saint Croix, was not available as a consquence of hurricane
damage.  
Two epochs of observations at 43\,GHz are presented in this work to study the
evolution of the millimeter-VLBI jet in total flux density through the radio
outburst and obtain a high-resolution view of the inner-jet
structure. The Brewster antenna could not be pointed and was excluded
from the first of these observations (on 2017-11-10), thereby resulting in a
significant loss of resolution particularly in the north-south
direction (see Fig.~\ref{fig:images}). The latter observation, on 2018-05-04, was made
with the full VLBA.
Data were recorded at a data bit rate of 2048\,Gb\,s$^{-1}$.  See Table \ref{table:obs}.

\begin{table}[htb]
\caption{VLBA observational journal experiments (\texttt{BR224})
}
\centering
\resizebox{\columnwidth}{!}{
\begin{tabular}{@{}lcr@{$\times$}lcc@{\,\,}c@{\,\,}c@{}}

\hline
\noalign{\smallskip}
Date & 
Freq. & \multicolumn{3}{c}{43\,GHz Beam} & $S_\mathrm{tot}$ & $S_\mathrm{peak}$ & $S_\mathrm{min}$ \\ 
\textsc{yyyy-mm-dd} & 
                      [GHz] & [{\tiny $\mu$as}&{\tiny $\mu$as}] & [$^\circ$] & [{\tiny mJy}] & [{\tiny mJy/beam}] & [{\tiny mJy/beam}] \\
\noalign{\smallskip}
\hline
\noalign{\smallskip}
2017-11-10$^\mathrm{a}$ &  
             15/23/43/86 &  1090&195 & $-19$  &  496 & 320 & 0.63 \\
\hline
2018-05-04 &  
             43/86 & 451&207 & $-7$ & 719 & 403 & 0.73 \\
\noalign{\smallskip}
\hline
\multicolumn{7}{@{}l}{\tiny $^\mathrm{a}$ Brewster and Saint Croix did not participate in this epoch.} \\
\end{tabular}
}
\label{table:obs}
\end{table}

We reduced the data following standard procedures. A full discussion of all observational data including
the polarization analysis and a multi-frequency analysis, and the discussion of the 86\,GHz data that suffered severe gain and performance variations, will be presented elsewhere.
In this work we focus on the imaging results at 43\,GHz
($\lambda$7\,mm) on November 10, 2017 and May 4, 2018, to study the evolution of
the millimeter-VLBI jet in total flux density through the radio outburst in order to obtain a high-resolution view of the inner-jet structure.  


\section{Results}


%

\begin{figure*}[htb]
\centering
\parbox[b]{.70\textwidth}{
\includegraphics[trim=70 70 60 85,clip, width=0.34\textwidth]{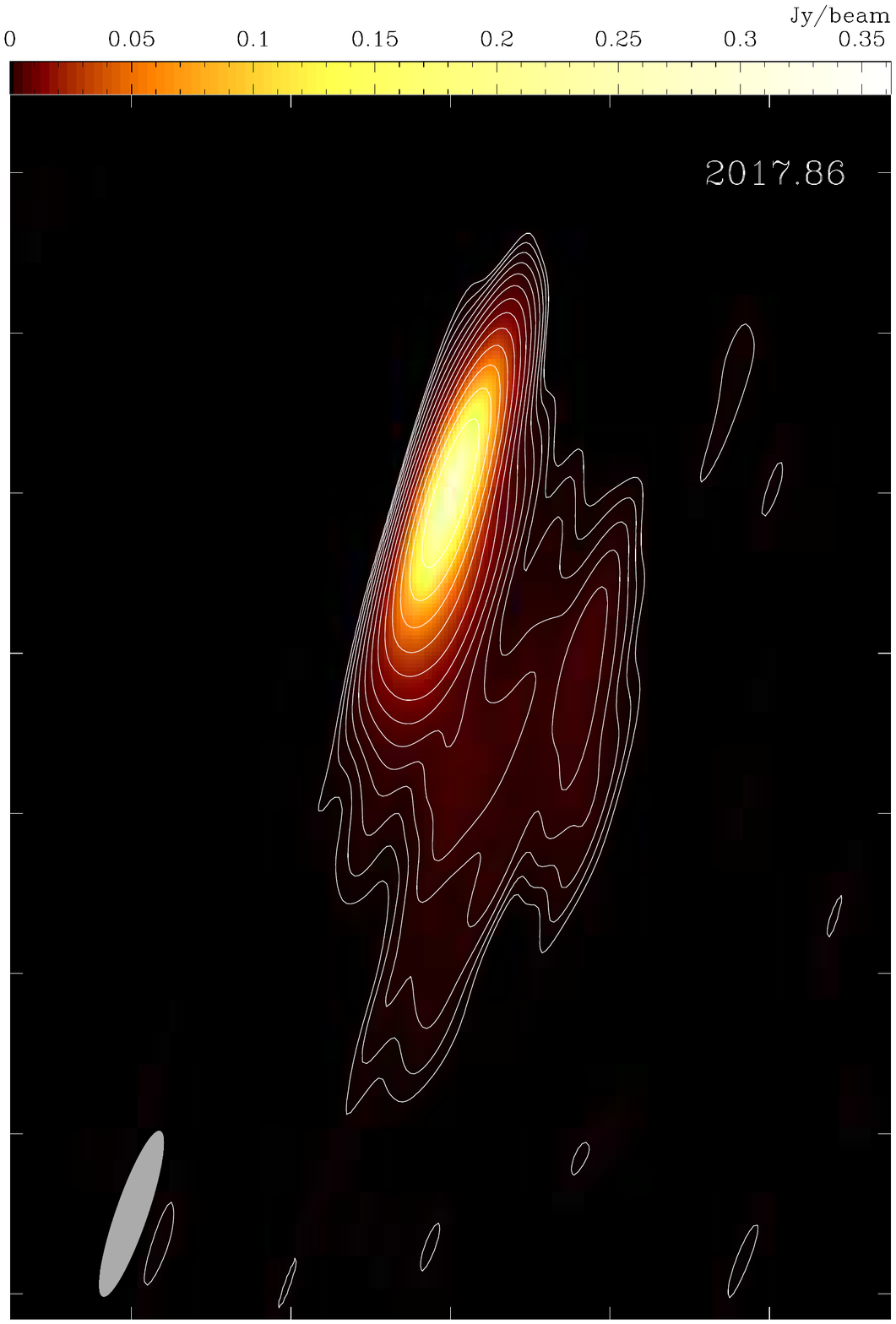}\hfill
\includegraphics[trim=70 70 60 85,clip, width=0.34\textwidth]{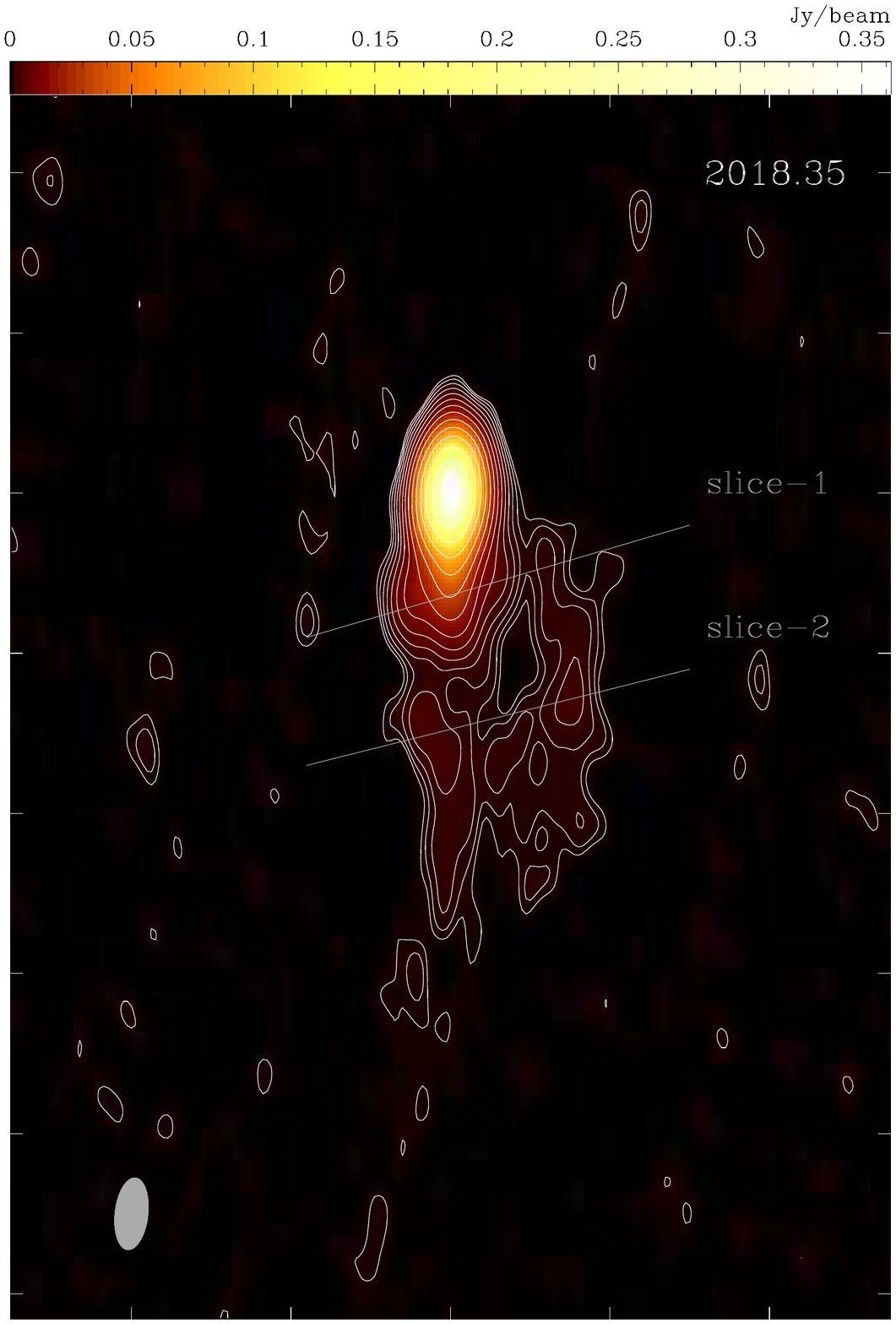}
}\hfill
\parbox[b]{.28\textwidth}{
\includegraphics[clip,width=0.25\textwidth]{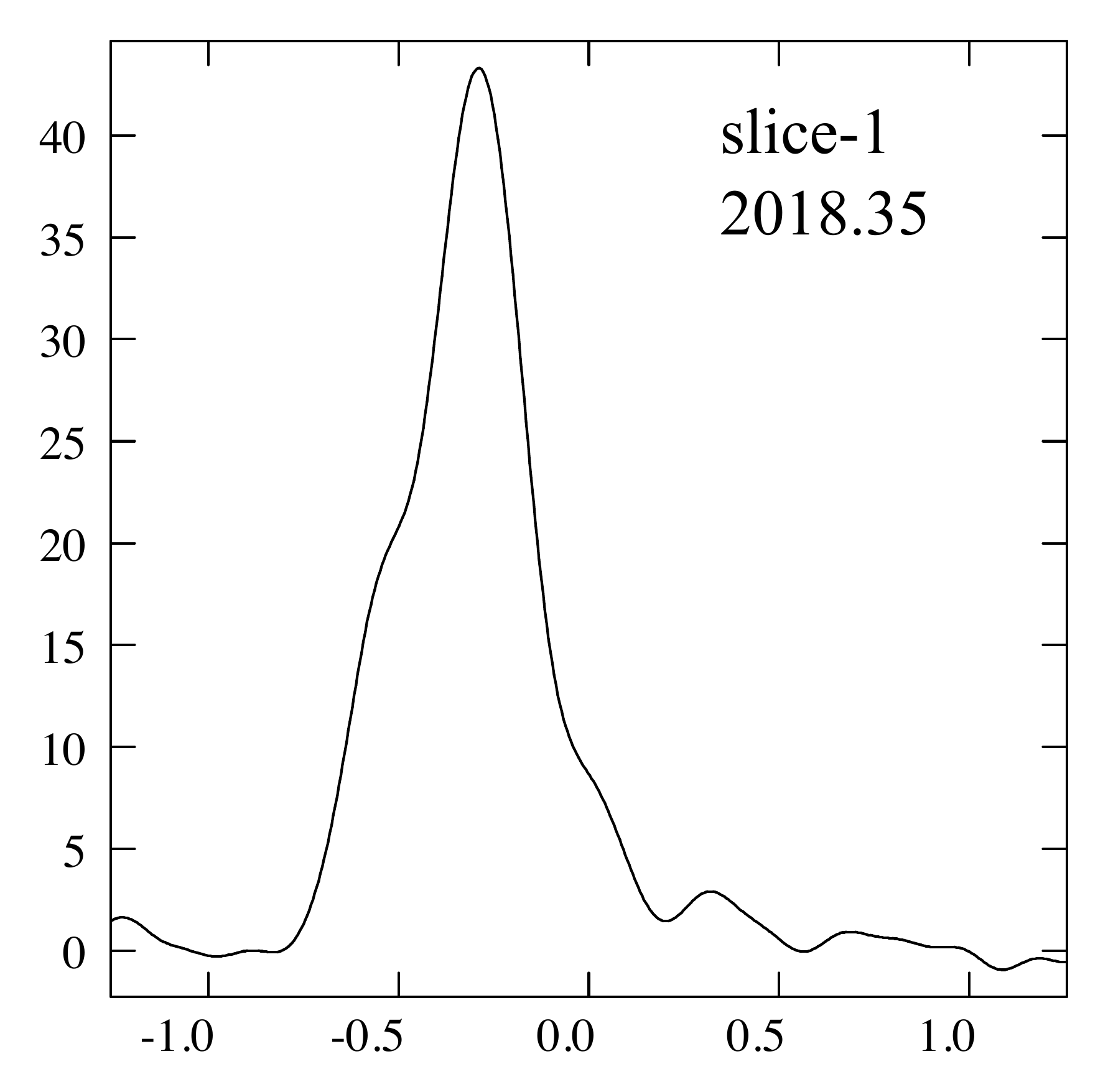}
\includegraphics[clip,width=0.25\textwidth]{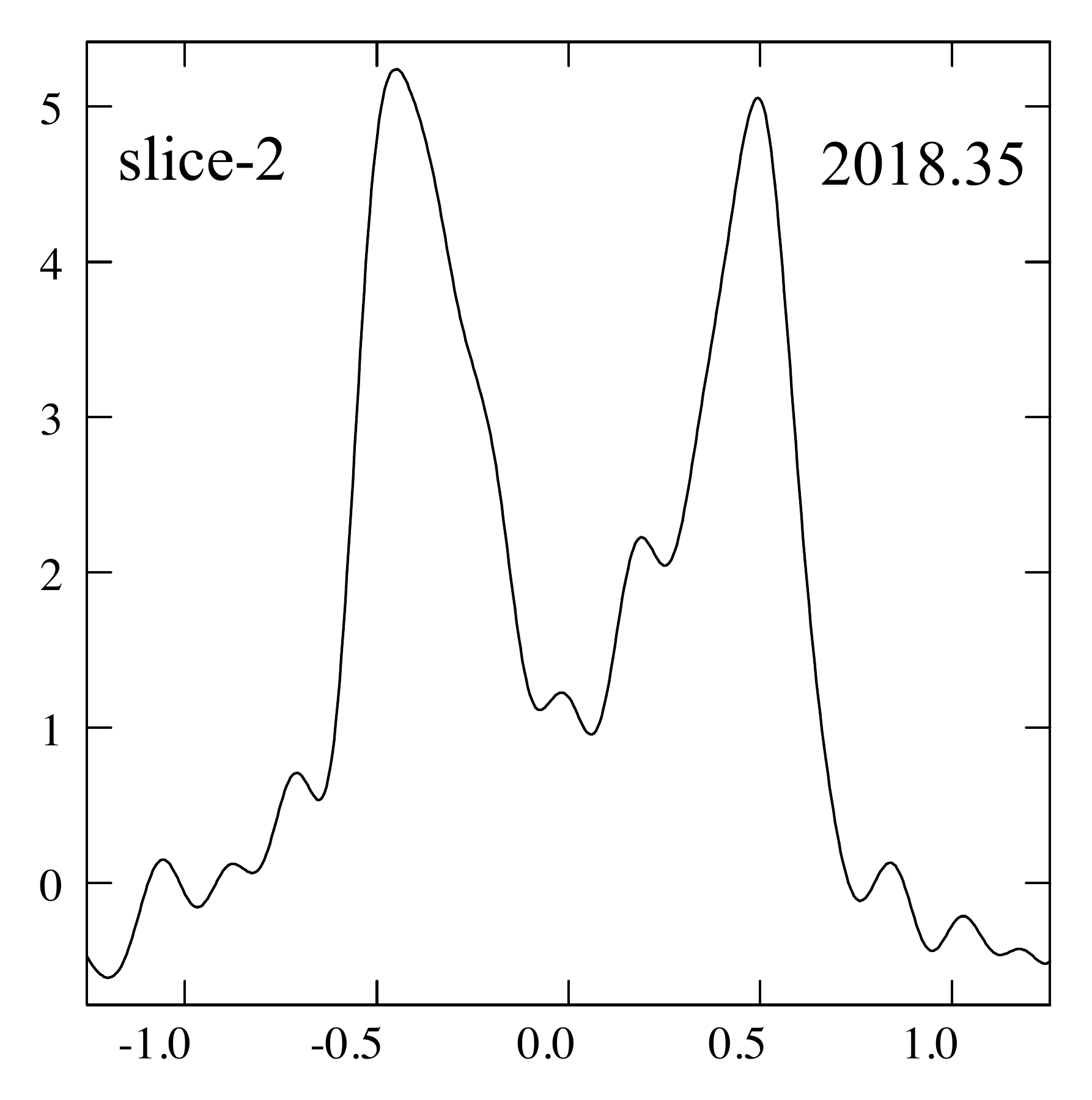}
}
\caption{Forty-three\,GHz VLBA images of \txs0506 on Nov 11, 2017 (left panel), and May 4, 2018 (middle panel). Apart from the different beam sizes, the resolved structure is not significantly different. Image parameters are given in Table 1.
The distance between the ticks on the axes is 1\,mas, corresponding to 4.78\,pc at the distance of \txs0506.  Right panels: Surface brightness profiles (mJy beam$^{-1}$ vs. mas, relative to the slice mid-point) for two slices transverse to the jet, as shown in the middle panel. \label{fig:images} }
\end{figure*}

\begin{figure*}[htb]
\centering
\parbox[b]{.70\textwidth}{
\includegraphics[trim=70 170 60 210,clip, width=0.34\textwidth]{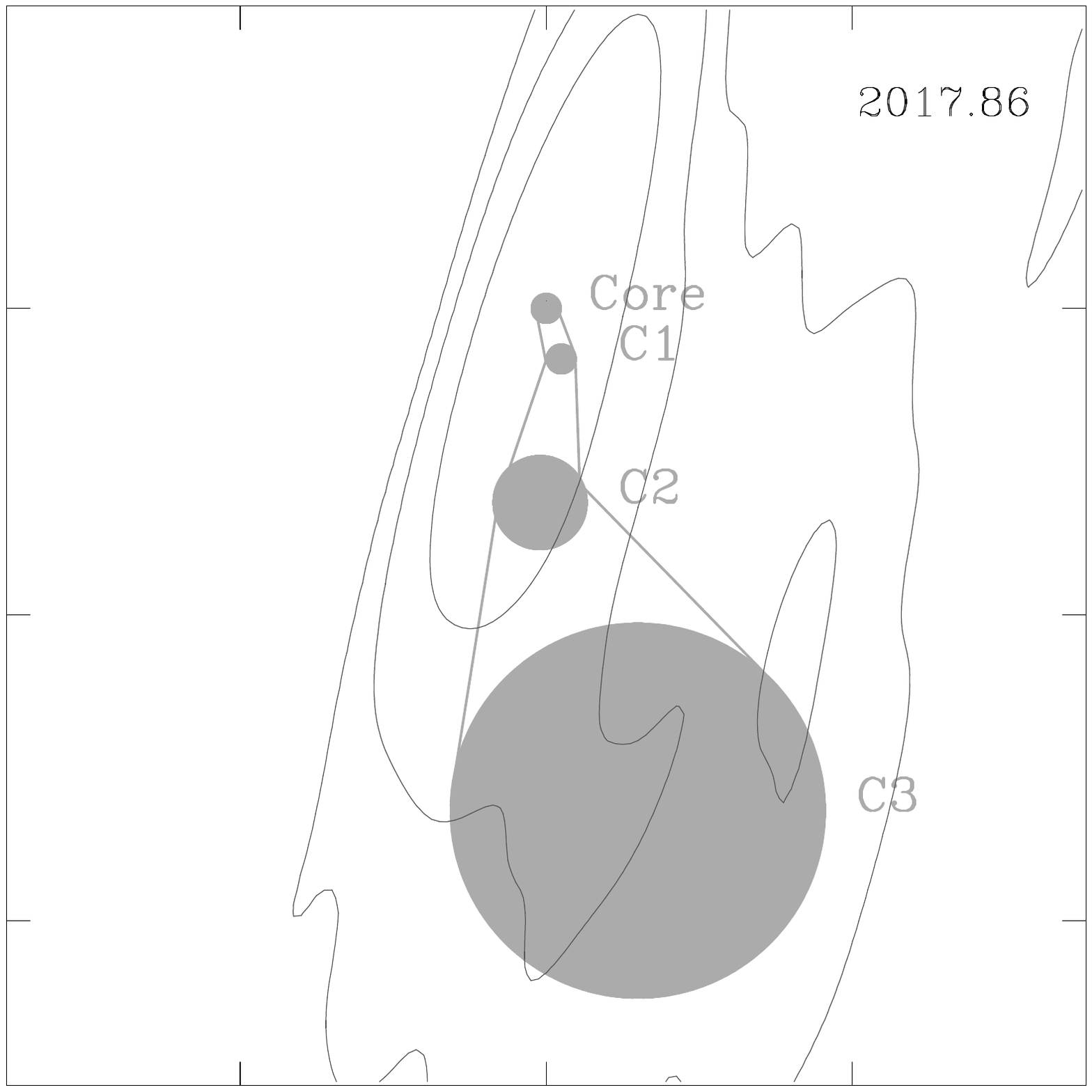}\hfill
\includegraphics[trim=70 170 60 210,clip, width=0.34\textwidth]{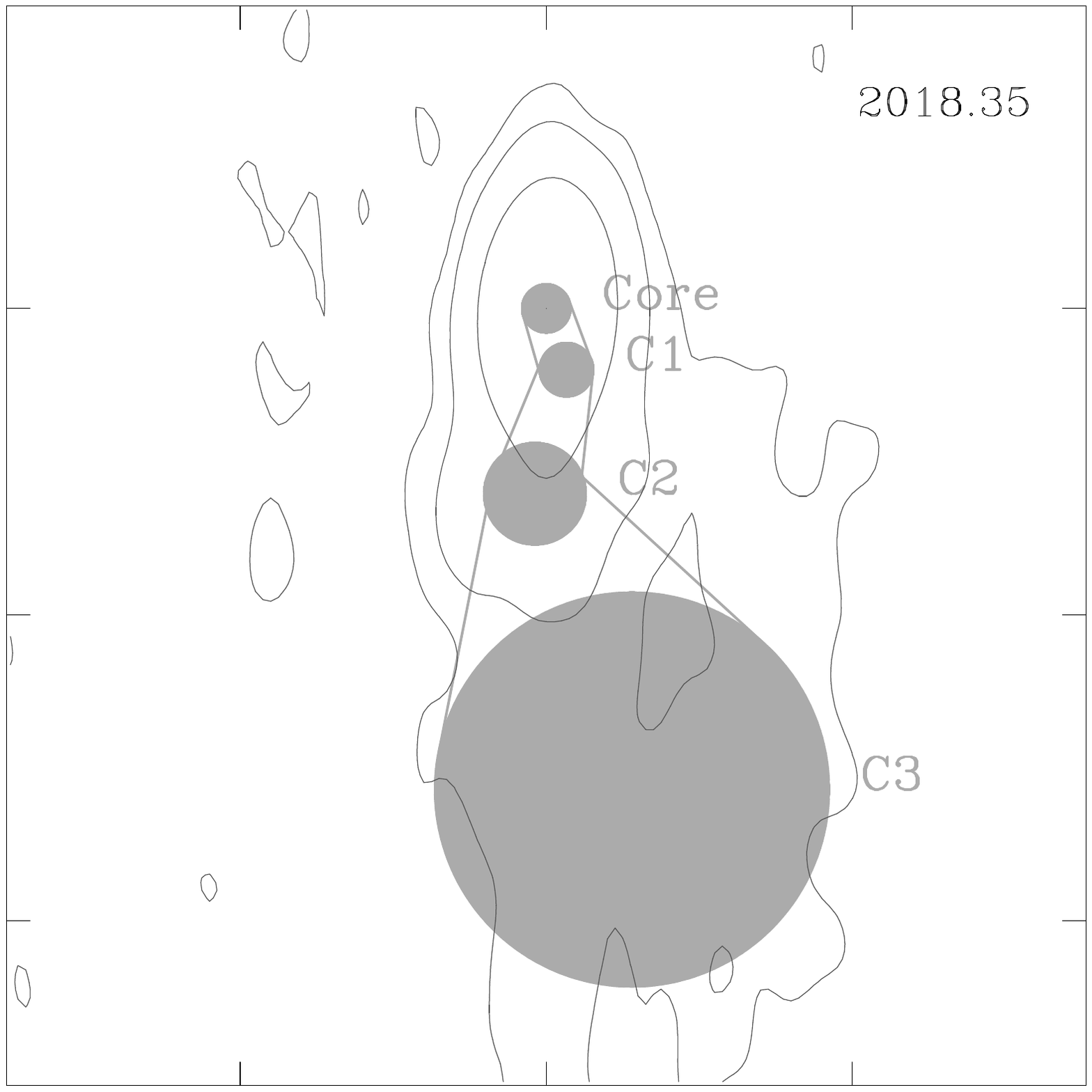}
}\hfill
\parbox[b]{.28\textwidth}{
\caption{Zoom into the 43\,GHz VLBA images of \txs0506 in Fig\,1, showing the Gaussian
model functions fit to the interferometric
visibilities and are presented in Table\,2.   \label{fig:modelfits} }
}
\end{figure*}



\begin{figure*}
\parbox[b]{.70\textwidth}{
\includegraphics[width=0.34\textwidth]{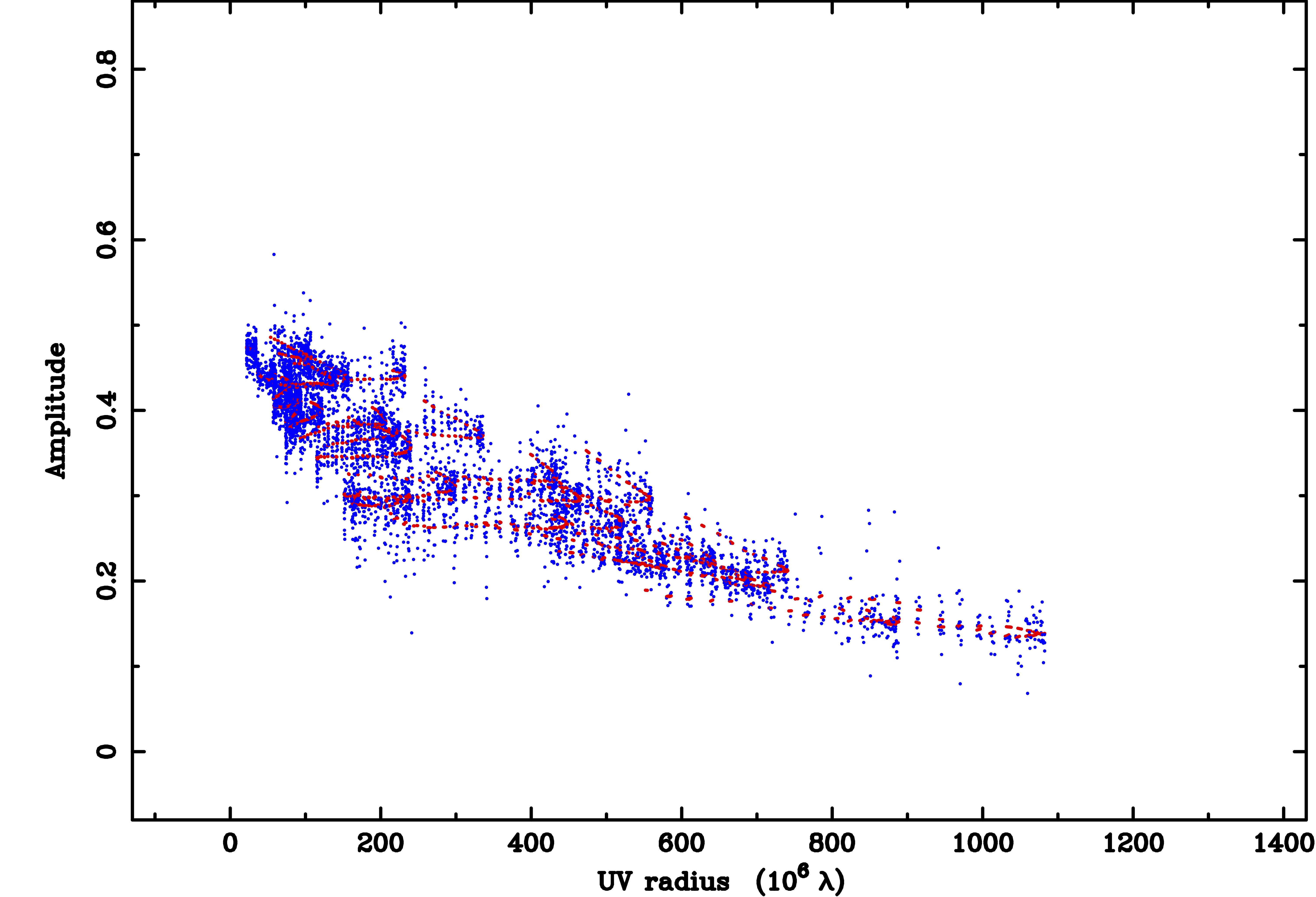}\hfill
\includegraphics[width=0.335\textwidth]{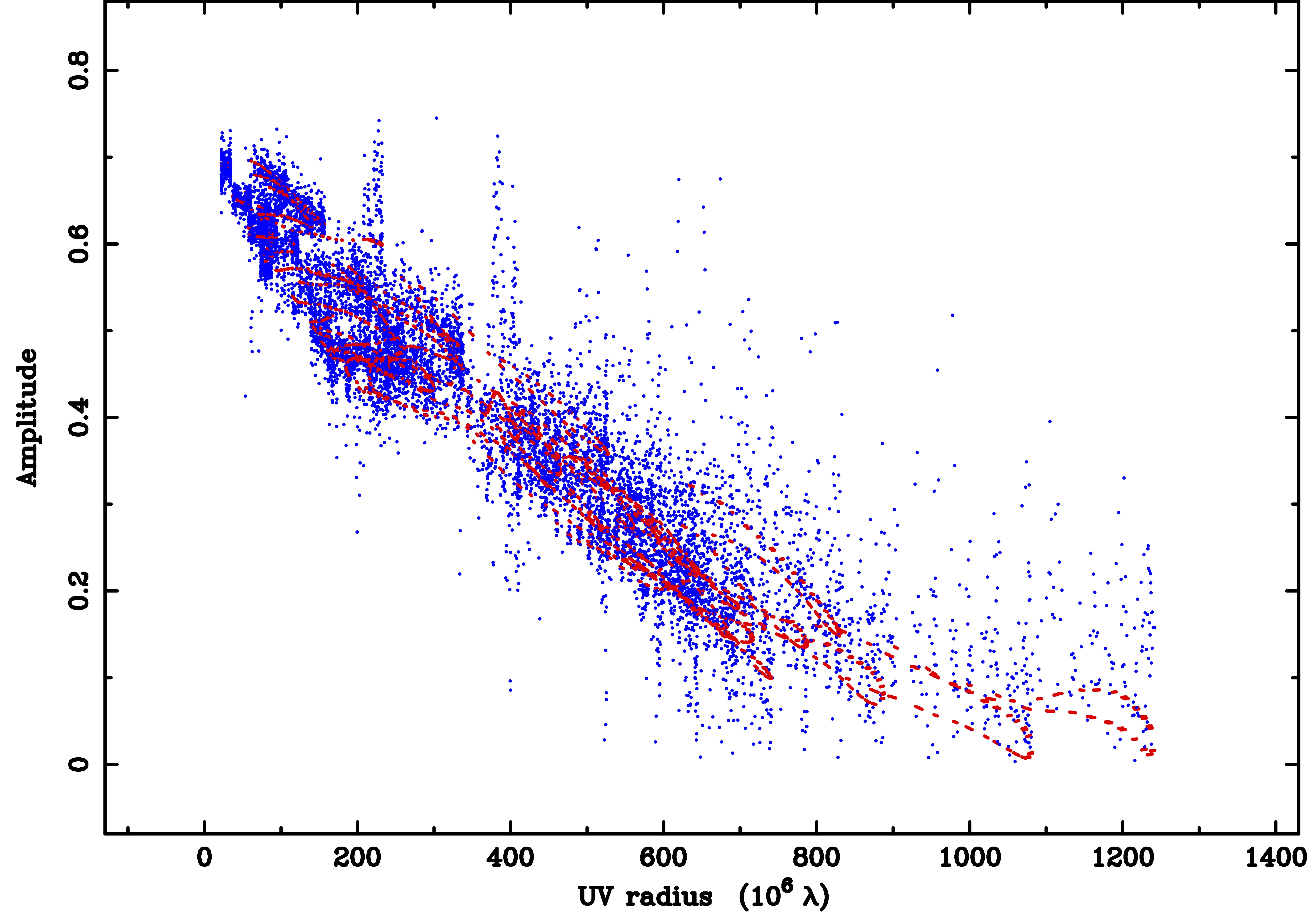}
}\hfill
\parbox[b]{.28\textwidth}{
\caption{Visibility amplitude (in Jy) as function of $(u,v)$-distance (in G$\lambda$) for Nov 11, 2017, and May 4, 2018. Data (averaged every 4\,min for clarity) are shown in blue. The \textsc{clean} model is shown in red. The data clearly show a brighter and less compact structure in May 2018.
\label{fig:radplot}
}
}
\end{figure*}

The $\lambda$7\,mm images presented 
in Fig.\,\ref{fig:images} show
a core-jet structure pointing to the south, which is similar
to but better resolved than the images of the MOJAVE survey at
$\lambda$2\,cm.  The image fidelity and particularly the angular resolution are different owing to the different observing conditions and array configurations.  A close morphological inspection, however, shows no significant structural differences on scales larger than about 1\,mas. On these scales, the jet does not show any clear isolated knots of locally enhanced brightness temperature for which speed measurements can be obtained. In particular, we can exclude the presence of a secondary jet core in the region roughly 1.2\,mas southwest of the primary core, as suggested by Britzen et al.~(\cite{bri19}), down to a limit of about 1\,mJy; this would correspond to 5$\sigma$ in our images.

The images of \txs0506\ show a compact core and a highly collimated inner jet within the inner parsec. The jet downstream shows a subsequently wider opening angle and the morphology is clearly limb brightened, reminiscent of the VLBI morphology found in several other teraelectronvolt blazars (e.g., Giroletti et al. \cite{giroletti04,giroletti08}; Piner \& Edwards \cite{piner14,piner18}). 
{We highlight the change in transverse structure from the inner parsec to the outer part of the jet by showing brightness profiles across the jet at two different locations (Fig.~\ref{fig:images}, right panels).}

We fit the jet structure using Gaussian components (illustrated in Fig.\,\ref{fig:modelfits}). Variability of the source is detected on scales smaller than about 1\,mas as illustrated in Fig.\,\ref{fig:radplot} (in which we plot visibility amplitude as a function of baseline length).
The source gets
brighter between November 2017 and May 2018, and the dominant emission region (the core) clearly expands, as evidenced from the fall
in the visibility function in the May 2018 epoch. 

\begin{table*}[htb]
\centering
\caption{Gaussian model fitting results.  \label{table:modelfit}}
\begin{tabular}{@{}c@{\,\,}c@{\,\,}c@{\,\,}c@{\,\,}c@{\,\,}c@{\,\,}c@{\,\,}c@{\hspace*{0.5cm}}c@{\,\,}c@{\,\,}c@{\,\,}c@{\,\,}c@{\,\,}c@{\,\,}c@{\,\,}c@{}}
\hline
\noalign{\smallskip}
Date & ID & $S$ & \multicolumn{1}{c}{$\Delta\alpha$} & $\Delta\delta$ & FWHM & $\vartheta_\textrm{app}$& $T_b$ & Date & ID & $S$ & \multicolumn{1}{c}{$\Delta\alpha$} & $\Delta\delta$ & FWHM & $\vartheta_\textrm{app}$ & $T_b$\\
\textsc{yyyy-mm-dd} &  & [mJy] & [$\mu$as] & [$\mu$as] & [$\mu$as] & $[^\circ]$& [K] & \textsc{yyyy-mm-dd} &  & [mJy] & [$\mu$as] & [$\mu$as] & [$\mu$as] & $[^\circ]$& [K]\\
\noalign{\smallskip}
\hline
\noalign{\smallskip}
2017-11-10 & Core & \textbf{275} &     -- &      -- &   \textbf{68}$^\mathrm{a}$ & --& $5.3\times10^{10}$ & 2018-05-04 & Core & \textbf{508} &     -- &      -- &  \textbf{158}$^\mathrm{a}$ & -- & $1.8\times10^{10}$\\
           & C1   & 110 &  $-49$ &  $-165$ &   94 & \textbf{4.3} & $1.1\times10^{10}$ &            & C1   &  91 &  $-66$ &  $-200$ &  175 & \textbf{2.4} & $2.6\times10^9$\\
           & C2   &  85 &   $20$ &  $-634$ &  304 & \textbf{12.7} & $7.2\times10^8$ &            & C2   &  77 & $37$ &  $-605$ &  332 & \textbf{10.7} & $6.2\times10^8$\\
           & C3   &  38 & $-299$ & $-1640$ & 1221 & \textbf{23.5} & $2.3\times10^7$ &            & C3   &  54 & $-280$ & $-1571$ & 1286 & \textbf{25.1} & $2.9\times10^7$\\
\noalign{\smallskip}
\hline
\multicolumn{16}{@{}l@{}}{\tiny {Note: $S$: total flux density; $\Delta\alpha$ and $\Delta\delta$: offsets in RA and DEC; FWHM: Gaussian FWHM, $\vartheta_\textrm{app}$: apparent opening viewing angle w.r.t. previous}}\\[-1pt]
\multicolumn{16}{@{}l@{}}{\tiny {feature; and $T_b$ brightness temperature. In 
boldface, relevant values in the discussion.  $^\mathrm{a}$: Formal fit uncertainties are 1.1\,$\mu$as and 0.2\,$\mu$as, respectively.}} \\[-1pt]
\end{tabular}
\end{table*}

We modeled the interferometric visibilities to
parametrize the source structure with Gaussian 
functions using the Levenberg-Marquardt algorithm 
implemented in the routine \textsc{modelfit}
in the software \textsc{Difmap} (Sheperd \cite{she97}).  
The results of this modeling are shown in
Table~\ref{table:modelfit} and Fig.\,\ref{fig:modelfits}.
A structure with four components 
aligned toward the south is consistently found in both
observing epochs, and opens rapidly in size; compare the full width half
maximum (FWHM) values in columns 6 and 13.  
{We note that following Kovalev et al.\ (\cite{kov05}, see their eq.\ 2) the minimum resolvable size of
a Gaussian component fitted to the visibilities would be of 21\,$\mu$as and 12\,$\mu$as for the first and the second epoch images, respectively, showing the robustness of our size estimates.}
We label these components Core (innermost, brightest component, in the millimeter regime often associated with a standing shock in the jet), as well as the jet features
C1, C2, and C3. 
C1 represents a bright compact knot of emission that partially blends with the core and is located just (0.17--0.20)\,mas to the south. 
C2 and C3, in contrast, represent larger diffuse emission regions at about 0.6\,mas and 1.7\,mas south and south-southwest of the core, respectively. 
The most external region corresponding to C3 can also be modeled with two Gaussian functions at the jet edges, providing similar model statistics as presented for the single-component features.  
The apparent opening angle $\vartheta_\textrm{app}$ between adjacent components increases monotonically with increasing distance from the core: the inner jet between the core and C1 appears highly collimated\footnote{
We tested the residual visibilities when subtracting the core component in both epochs to confirm that the core 
is clearly resolved. In the case of the May 2018 epoch, this is immediately apparent from inspection of the overall visibility amplitudes as a function of baseline length (see Fig.~\ref{fig:radplot}).}  with an apparent opening angle of only $2^\circ$ to $4^\circ$; the angle between C1 and C2 is already substantially larger at  $11^\circ$ to $13^\circ$, whereas the outermost region between C2 and C3 has an apparent opening angle of $23^\circ$ to $25^\circ$, which is compatible with the value of $28^\circ$ determined at 15\,GHz from stacked MOJAVE images in Pushkarev et al.~(\cite{pushkarev17}).

The flux density of the core changes between both epochs,
almost doubling its value within six months while the other jet features do not show strong variability. 
Thus, we can attribute the radio outburst, seen at all radio frequencies, to a region inside the 7\,mm VLBI core.
Remarkably, the core size
grows with
time, so that the brightness temperature associated with it
drops by a factor of three between the two epochs despite the rising flux density
(compare columns 7 and 14). 
The outer rim of the core component moves by {$(158-68)/2=45$\,$\mu$as (with a formal uncertainty smaller than 1\,$\mu$as)}
within six months, which
at a scale of 4.78\,pc/mas, corresponds to a speed of about two times the speed of light. This effect of apparent superluminal expansion can be understood as a projection effect that occurs when relativistic plasma moves at a small angle to the line of sight, as is commonly observed for isolated jet knots in blazars (see, e.g., Cohen et al.~\cite{cohen2007}).  Superluminal expansion in the core of the Seyfert galaxy \object{III\,Zw\,2} was reported in Brunthaler et al.~(\cite{bru00}).
An apparent superluminal core expansion is more unusual, in particular because the maximal speed of jet features in this object is below 1\,c (Lister et al.~\cite{lis19}).

\section{Discussion}
\subsection{Structured jets as neutrino emitters}

Bright neutrino emission is generally expected to be produced in flat-spectrum radio quasars (FSRQs) owing to the presence of a strong optical/UV seed photon field allowing interactions with high-energy protons inside the relativistic jet and subsequent pion production and decay (see, e.g., Murase et al.~\cite{murase14}). Lacking a strong broad-line region, BL\,Lacs, on the other hand, have been considered to be inefficient in neutrino production unless their jets carry an unexpectedly high power in accelerated protons (see Cerruti et al.~\cite{cerutti18}). In this context, the classification of \txs0506\ has created severe theoretical challenges for the 
neutrino emission models of this source. As a way out of this gridlock, it has been proposed that another external photon field may provide the seed photons for the photopion production. Popular variants of this approach are models that assume a spine-sheath structure of the jet with a highly relativistic inner spine and a slower mildly to non-relativistic outer sheath (Ghisellini et al.~\cite{ghisellini05}). As a result of the strong velocity difference, the copious photons emitted in the slower sheath appear substantially boosted in the rest frame of the relativistic spine, so that enhanced photopion production can occur. Consequently, such a stratified jet is expected to be a much brighter neutrino source than a uniform (one-zone) jet (Tavecchio et al.~\cite{tav14}). At the same time, this model can solve the Doppler problem of the apparent contradiction of a low Doppler factor suggested by low apparent VLBI speeds of the jet and much higher Doppler factors implied by $\gamma$-ray observations of teraelectronvolt blazars (see, e.g., Blasi et al. \cite{blasi2013}, Piner \& Edwards \cite{piner18}). Observational evidence for such a spine-sheath configuration of jets affected by the Doppler problem can in principle come from VLBI observations (e.g., Giroletti et al. \cite{giroletti04,giroletti08}). 

\subsection{Increasing opening angle and superluminal core expansion}

Both the increase of the apparent opening angle with distance and the rapid core expansion with time can be interpreted within two scenarios: first, the jet deceleration, as suggested by Georganopoulos \& Kazanas (\cite{geo03}); and second, a spine-sheath structure of the jet, as suggested by Ghisellini et al.~(\cite{ghisellini05}).
The inner jet between the core and C1 clearly appears to be highly collimated, which is suggestive of a fast relativistic flow. Beyond C1, the jet might decelerate and bend first to the east (C2) and then back southward (C3). This would be consistent with the model suggested by Georganopoulos \& Kazanas (\cite{geo03}). Alternatively, the two components C2 and C3 might represent local brightness enhancements within a much wider diffuse jet. 
This second interpretation is consistent with the scenario put forward by Ghisellini et al.~(\cite{ghisellini05}).

The apparent superluminal expansion of the core can be interpreted as an adiabatic expansion of a highly relativistic plasma cloud traveling down the jet spine with an opening angle $\phi$ at an inclination angle $\vartheta$ to the line of sight. If we assume that $\vartheta < \phi$, the apparent core expansion would correspond to a component traveling down the jet with apparent $(28-57)$\,c for a range of opening angles between $4^\circ$ and  $2^\circ$.  In this case, we are looking into the jet and the external Compton emission (from a seed photon field originating in the sheath) dominates over the synchrotron self Compton (SSC) emission and the emission from hadronic processes. This is supported by the apparent opening angle of the inner jet between the core and component C1 of only $(2-4)^\circ$. These values are fully consistent with the assumptions made in the  modeling of the SED of \txs0506\ made by Ansoldi et al.~(\cite{magic2018}). Recently, Zhang et al. (\cite{zhang19}) have presented a model, which explains both the 2014--2015 ``neutrino flare'' (IceCube Collaboration \cite{ice18b}) and the detection of the single IceCube-170922A high-energy neutrino in 2017, by involving a persistent external photon field such as an outer jet sheath.

The expansion and deceleration of FRI jets at hundreds of parsecs to kiloparsecs (e.g., Laing \& Bridle \cite{lb14}) has been interpreted by several authors via collective interactions between jets and stellar winds, clouds, or even supernovae (Hubbard \& Blackman \cite{hb06}; Barkov, Aharonian \& Bosch-Ramon \cite{bab10}; Bosch-Ramon, Perucho \& Barkov \cite{br+12}; Perucho et al.~\cite{pe14}; Vieyro, Bosch-Ramon \& Torres-Alb\`a~\cite{vie19}). The fast jet expansion observed in \txs0506\ is reminiscent of such ``geometrical flaring'', although at parsec scales in this case ($r\leq 15$\,pc to $r\sim 30$\,pc for a viewing angle between $4^\circ$ and $2^\circ$), which could be explained in terms of a single collision with a massive star (e.g., Hubbard \& Blackman \cite{hb06}).
Evidence for such direct jet-star interactions has indeed been found, for example, by Müller et al. (\cite{mue14}). Interestingly, these scenarios naturally embed a population of protons in the jet, thus facilitating hadronic processes and neutrino production. These interactions are more probable toward the galactic nucleus, where the densities of 
stars and clouds are larger, and is precisely the region in which we observe the geometrical flaring to begin.
Indeed, the mass loading from star-jet interactions in AGN has been considered a substantial contribution to the high-energy emission of blazars (De la Cita et al. \cite{dlc16}, Perucho et al. \cite{pe17}, Torres-Alb\`a \& Bosch-Ramon \cite{tab19}, Vieyro, Bosch-Ramon \& Torres-Alb\`a \cite{vie19}). In particular the SED and neutrino emission of \txs0506\ has also been interpreted and modeled in this context (Sahakyan \cite{sah18}, Wang et al. \cite{wan18}).

\subsection{Discussion of other neutrino-source candidates}

The FSRQ 
\object{PKS\,1424$-$418}
has been associated with a petaelectronvolt neutrino with a chance coincidence of about $2 \sigma$ (Kadler et al.~\cite{kad16}) based on the occurrence of a major radio outburst and a high-fluence kiloelectronvolt-gigaelectronvolt outburst coincident
with the neutrino event IC35 in position and time.  The {Tracking Active Galactic Nuclei with Austral Milliarcsecond Interferometry (TANAMI}) 8.6\,GHz observations
reveal a substantial brightening of the VLBI core on scales smaller than 1\,mas. The high brightness temperature, flat radio spectrum, and rapid increase in flux density suggest a highly relativistic jet on these sub-milliarcsecond scales. In the context of our new finding regarding the spine-sheath structure of the \txs0506\ jet, it is interesting to note that the jet of \object{PKS\,1424$-$418} on scales of dozens of milliarcseconds has been found to be very diffuse and resolved, as pointed out by Ojha et al.~(\cite{ojha10}), and shows one of the widest opening angles in the whole TANAMI sample. This morphology is consistent with the morphology displayed by \txs0506\ as presented above. It should be noted, however, that several other VLBI jets have been observed to display similar wide-opening jet morphologies, yet these jets have not been associated with high-confidence neutrino events. A spine-sheath or geometrically flaring VLBI morphology alone is clearly not a sufficient condition to predict neutrino emission.

\section{Summary and conclusions}

We imaged the parsec-scale morphology of {the jet} in \txs0506\  at $\lambda$7\,mm.  The source shows a compact high brightness-temperature core and a highly collimated inner jet within the inner parsec from the core. Further downstream, the morphology changes into a subsequently limb-brightened jet with a wider opening.

We identified the location of the radio outburst associated with the IceCube-170922A neutrino event to within the core component, i.e., to linear deprojected scales smaller than 30\,pc (unless the angle to the line of sight is much smaller than $2^\circ$). 

The core expands with apparent superluminal velocity within eight months after the detection of the IceCube neutrino. This, along with the high brightness temperature of the core, suggests the presence of a highly relativistic beam on these scales.

The source morphology beyond 1\,mas from the core is suggestive of a slower flow, either due to overall deceleration or  jet-transversal velocity stratification. The slower flow can serve as a source of seed photons for photopion production and subsequent neutrino emission to explain the IceCube-170922A event as modeled, for example, by Ansoldi et al.~(\cite{magic2018}).

Further VLBI analysis of this source will include the study of the
polarization properties and the spatially resolved milliarcsecond-scale continuum spectrum, but this is 
beyond the scope of the present publication.

\begin{acknowledgements}
We are especially thankful to N.\ MacDonald for valuable comments. 
We also acknowledge C.\ Breu for support during the data calibration. 
We thank the anonymous referee for the rapid and very constructive comments to the manuscript.
The Very Long Baseline Array is a facility of the National Science Foundation operated under cooperative agreement by Associated Universities, Inc. 
F.\,K.\ was supported as an Eberly Research Fellow by the Eberly College of Science at the Pennsylvania State University.
This research has made use of National Aeronautics and Space Administration's (NASA) Astrophysics Data System, 
and of the NASA/IPAC Extragalactic Database (NED)
which is operated by the Jet Propulsion Laboratory, 
California Institute of Technology, under contract with NASA.
\end{acknowledgements}

%
%



\end{document}